\documentclass[english,a4paper,manuscript]{revtex4}
\usepackage[T1]{fontenc}
\usepackage[latin1]{inputenc}

\makeatletter

\providecommand{\tabularnewline}{\\}

\makeatletter

\makeatletter

\usepackage{geometry}

\makeatother

\makeatother

\makeatother

\usepackage{babel}

\begin{document}

\title{The Rayleigh-Brillouin Spectrum in Special Relativistic Hydrodynamics}

\author{A. L. Garcia-Perciante$^{1}$, L. S. Garcia-Colin$^{2}$, A. Sandoval-Villalbazo$^{3}$}

\address{$^{1}$Depto. de Matematicas Aplicadas y Sistemas, Universidad Autonoma
Metropolitana-Cuajimalpa, Artificios 40 Mexico D.F 01120, Mexico.\\
 $^{2}$Depto. de Fisica, Universidad Autonoma Metropolitana-Iztapalapa,
Av. Purisima y Michoacan S/N, Mexico D. F. 09340, Mexico. Also at
El Colegio Nacional, Luis Gonzalez Obregon 23, Centro Historico, Mexico
D. F. 06020, Mexico. \\
 $^{3}$Depto. de Fisica y Matematicas, Universidad Iberoamericana,
Prolongacion Paseo de la Reforma 880, Mexico D. F. 01219, Mexico}

\address{}

\begin{abstract}
In this paper we calculate the Rayleigh-Brillouin spectrum for a relativistic
simple fluid according to three different versions available for a
relativistic approach to non-equilibrium thermodynamics. An outcome
of these calculations is that Eckart's version predicts that such
spectrum does not exist. This provides an argument to question its
validity. The remaining two results, which differ one from another,
do provide a finite form for such spectrum. This raises the rather
intriguing question as to which of the two theories is a better candidate
to be taken as a possible version of relativistic non-equilibrium
thermodynamics. The answer will clearly require deeper examination
of this problem. 
\end{abstract}
\maketitle

\section{Introduction}

It is a well known fact that light scattering by a simple fluid in
equilibrium at a certain temperature $T$ and pressure $p$ is one
conclusive test to verify Onsager's linear regression of fluctuations
hypothesis \citep{onsager}, a basic assumption in classical irreversible
thermodynamics \citep{dgm,meix}. Not only it constitutes the core
behind the proof of Onsager's reciprocity theorem but it also guarantees
that the equilibrium state of the fluid is stable under such fluctuations.
Thus, it is legitimate to ask if the various formulations of irreversible
relativistic thermodynamics so far available are at grips with such
hypothesis. Even in an indirect way, this would indicate that such
theories can be tested experimentally. Indeed, in the classical case
one measures the so-called dynamic structure factor of the fluid $\mathcal{S}\left(\vec{q},\,\omega\right)$
which represents the energy scattered by the fluid from an incoming
wave of wavelength $\lambda$, as a function of frequency. The outcome
of this measurement is the well known Rayleigh-Brillouin (RB) spectrum
\citep{mountain,berne,boon}. Its central peak, the Rayleigh peak,
has a width proportional to the thermal diffusivity $D_{T}=\kappa/\rho_{0}C_{V}$
where $\kappa$ is the thermal conductivity, $\rho_{0}$ the equilibrium
density and $C_{V}$ the specific heat at constant volume. This peak
represents the intensity of the thermal (entropy) fluctuations. Symmetrically
located with respect to this peak there appear two peaks, the Brillouin
peaks, which represent the fluctuations arising from the mechanical
dissipative processes of the fluid, the sound or light absorption.
They are located at $\omega=\pm C_{0}k$ from the central peak, $C_{0}$
being the velocity of sound, if the probe is a sound wave. Their width
is given by the famous Stokes-Kirchhoff's formula namely,\begin{equation}
\Gamma=\frac{1}{2}\left\{ \frac{1}{\rho_{0}}\left(\frac{4}{3}\eta+\xi\right)+\frac{\gamma-1}{\gamma}D_{T}\right\} \label{eq:1}\end{equation}
 where $\gamma=C_{p}/C_{V}$, $\eta$ and $\xi$ being the shear and
bulk viscosities, respectively.

The important feature here is that the precise form of this spectrum
can be obtained by solving the linearized Navier-Stokes-Fourier equations
of hydrodynamics for the perturbations (or fluctuations) $\delta T$,
$\delta\rho$ and $\delta\vec{u}$ present in the fluid due to its
microscopic structure.

Now, it may be so that this experiment per se could not be easily
carried out in a laboratory in a relativistic regime for technological
reasons. However, one surely must expect that the corresponding linearized
equations of relativistic hydrodynamics lead to a relativistically
modified spectrum which reduces to its classical counterpart in the
non-relativistic limit. This is precisely the motivation of this paper.
We wish to calculate the RB spectrum for the linearized relativistic
hydrodynamic equations that arise in three cases: using the Eckart-Landau
Lifshitz formalism \citep{Eckart,L&L}, a relativistic generalization
of Meixner's theory \citep{JNT06} and considering the equations obtained
by the authors \citep{physica A,ERE09} when the acceleration term
in Eckart's theory is expressed in terms of $\nabla p$ using Euler's
equations. We will refer to this case as the modified Eckart's theory.

In Sects. II-V we shall establish the system of linearized relativistic
fluid equations for the three alternatives mentioned above and analyze
the modifications to the RB spectrum in each case. Section VI is devoted
to the discussion of the results and final remarks.

\section{MEIXNER-like formalism}

The first formalism we wish to analyze consists of the relativistic
generalization of Meixner's formalism \citep{JNT06}. In it, the heat
flux is not considered as part of the momentum-energy tensor but is
included in a separate total energy flux conservation equation. As
a result, the constitutive equation for the heat flux retains its
Fourier-type structure. The hydrodynamic equations for this formalism
have been obtained elsewhere \citep{ERE09,JNT06}. The fluctuations,
here denoted by a $\delta$ prefix, evolve to the equilibrium state
following the linearized version of such equations. Thus, the dynamics
of the fluctuations is given by

\begin{equation}
\delta\dot{n}+n_{0}\delta\theta=0\label{eq:m1}\end{equation}
 \begin{equation}
\tilde{\rho}_{0}\delta\dot{\theta}+\frac{1}{n_{0}\kappa_{T}}\nabla^{2}\delta n+\frac{\beta}{\kappa_{T}}\nabla^{2}\delta T-A\nabla^{2}\delta\theta=0\label{eq:m2}\end{equation}
 \begin{equation}
\delta\dot{T}+\frac{T_{0}\beta}{n_{0}c_{n}\kappa_{T}}\delta\theta-D_{T}\nabla^{2}\delta T=0\label{eq:m3}\end{equation}
 The second equation is a balance equation for the longitudinal component
of the fluctuations in the hydrodynamic velocity given by $\delta\theta\equiv\delta u_{;\nu}^{\nu},$
where $u^{\nu}$ is the hydrodynamic velocity four-vector. This equation
is obtained by calculating the divergence of the momentum balance
equation, a procedure which decouples the transverse mode whose dynamics
has been already analyzed in a separate work \citep{Nos1}. Here $n$
is the particle number density, $T$ the temperature, $\kappa_{T}$
the isothermal compressibility, $\beta$ the thermal expansion coefficient,
$C_{n}$ the heat capacity at constant particle density and $A=\zeta+4\eta/3$.
We have defined $\tilde{\rho}_{0}=\left(n_{0}\varepsilon_{0}+p_{0}\right)/c^{2}$
where $\varepsilon_{0}$ and $p_{0}$ are the internal energy and
pressure respectively. The naught subscripts denote equilibrium values,
the semi-colon a covariant derivative and a colon a component of a
gradient. Greek indices run from 1 to 4 and Latin ones from 1 to 3.

As mentioned above, the procedure to obtain the spectrum is the standard
one and involves calculating the dispersion relation arising from
the determinant of the Fourier-Laplace transformed hydrodynamic system
of equations. For Eqs. (\ref{eq:m1})-(\ref{eq:m3}) we obtain that

\begin{equation}
\left|\begin{array}{ccc}
s & n_{0} & 0\\
-\frac{1}{\tilde{\rho}_{0}n\kappa_{T}}q^{2} & s+\frac{A}{\tilde{\rho}_{0}}q^{2} & -\frac{\beta}{\tilde{\rho}_{0}\kappa_{T}}q^{2}\\
0 & \frac{T_{0}\beta}{n_{0}c_{n}\kappa_{T}} & s+D_{T}q^{2}\end{array}\right|=0\label{eq:m4}\end{equation}
 which yields a cubic dispersion relation which can be written as
follows\begin{equation}
s^{3}+a_{2}s^{2}q^{2}+s\left(a_{3}q^{4}+a_{4}q^{2}\right)+a_{5}q^{4}=0\label{eq:m5}\end{equation}
 where the coefficients are given by\[
a_{2}=\frac{A}{\tilde{\rho}_{0}}+D_{T}\]
 \begin{equation}
a_{3}=\frac{A}{\tilde{\rho}_{0}}D_{T}\label{eq:a3}\end{equation}
 \[
a_{4}=\frac{\gamma}{\kappa_{T}\tilde{\rho}_{0}}\]
 \[
a_{5}=\frac{D_{T}}{\kappa_{T}\tilde{\rho}_{0}}\]
 and we have used the relation $\frac{\beta^{2}T_{0}}{c_{n}n_{0}\kappa_{T}}=\frac{c_{p}-c_{n}}{c_{n}}=\gamma-1$.
One can easily show that Eq. (\ref{eq:m2}) has one real root given
by \begin{equation}
s_{1}=-\frac{a_{5}}{a_{4}}q^{2}\label{eq:sm}\end{equation}
 and a pair of conjugate roots\begin{equation}
s_{2,3}=\left(-\frac{a_{2}}{2}+\frac{a_{5}}{2a_{4}}\right)q^{2}\pm i\sqrt{a_{4}}q\label{eq:s23m}\end{equation}
 The analysis follows exactly as in the non-relativistic case \citep{mountain}-\citep{boon}
where it is shown that Eqs. (\ref{eq:sm}) and (\ref{eq:s23m}) are
valid up to terms of order $q^{4}$. Recalling that $S\left(\vec{q},\omega\right)$
is the density-density self correlation function, we may plot the
ratio between the dynamic and static structure factors, $S\left(\vec{q},\omega\right)/S\left(q\right)$
as a function of $\omega$ for a fixed $q$ which should yield three
peaks given by the roots of the cubic equation. The mean width of
the central peak, the Rayleigh peak, is determined by the real root
given in Eq. (\ref{eq:sm}) and thus, for the Meixner, case we obtain
a width\begin{equation}
\Delta_{RM}=\frac{D_{T}}{\gamma}q^{2}\label{eq:deltarm}\end{equation}
 A correction due to the modified value of the thermal conductivity
$\kappa$ for relativistic fluids will arise and thus one expects
to observe a change in the width of the peak.

The location and width of the other two peaks, the Brillouin peaks,
are determined by the conjugate roots given in Eq. (\ref{eq:s23m}).
The location of the peaks is given by the imaginary part while the
width is given by the real part. In this case the doublet appears
at $\omega=\pm\sqrt{a_{4}}q$ so that from Eqs. (\ref{eq:a3}) it
follows that \begin{equation}
\omega_{M}=\pm\sqrt{\frac{\gamma}{\kappa_{T}\tilde{\rho}_{0}}}q\label{eq:as}\end{equation}
 One should find a shift in this location due to the relativistic
value of $\tilde{\rho}_{0}$. The width of the doublet is given by
the real part of $s_{2,3}$, that is\begin{equation}
\Delta_{B}=\left(\frac{a_{2}}{2}-\frac{a_{5}}{2a_{4}}\right)q^{2}\label{eq:bes}\end{equation}
 and in this case we have\begin{equation}
\Delta_{BM}=\frac{1}{2}\left(\frac{A}{\tilde{\rho}_{0}}+\frac{\gamma-1}{\gamma}D_{T}\right)q^{2}\label{eq:ces}\end{equation}
 where once again, a correction due to the relativistic values of
$D_{T}$ and $\tilde{\rho}_{0}$ is expected. In both Eqs. (\ref{eq:deltarm})
and (\ref{eq:ces}) one recovers the non-relativistic expressions
when $c\rightarrow\infty$ as in this case $\tilde{\rho}_{0}\rightarrow\rho_{0}$,
the equilibrium density.

\section{Eckart's framework}

Eckart's theory for relativistic fluids \citep{Eckart} is based on
the construction of an momentum-energy tensor where heat flux is included.
As a consequence, in order to satisfy the second law of thermodynamics,
he proposed a rather controversial constitutive equation for the heat
flux in which a hydrodynamic acceleration term is included. This proposal
has been claimed to render the theory unphysical and motivated the
use of extended theories as alternatives \citep{israel}-\citep{HL},
as has been thoroughly discussed \citep{physica A,ERE09,Nos1}. The
aim of this section of to present the effect of such constitutive
equation in the structure of the RB spectrum. Thus, once again, the
starting point is the linearized set of equations for the fluctuations
in a relativistic fluid, now within Eckart's theory, which are shown
in Ref. \citep{Nos1} to read as\begin{equation}
\delta\dot{n}+n_{0}\delta\theta=0\label{eq:2}\end{equation}
 \begin{eqnarray}
\tilde{\rho}_{0}\delta\dot{\theta}+\frac{1}{n_{0}\kappa_{T}}\nabla^{2}\delta n+\frac{\beta}{\kappa_{T}}\nabla^{2}\delta T\nonumber \\
-A\nabla^{2}\delta\theta-\frac{\kappa}{c^{2}}\nabla^{2}\delta\dot{T}-\frac{\kappa T_{0}}{c^{4}}\delta\ddot{\theta} & =0\label{eq:3}\end{eqnarray}
 \begin{equation}
\delta\dot{T}+\frac{T_{0}\beta}{n_{0}c_{n}\kappa_{T}}\delta\theta-D_{T}\nabla^{2}\delta T-\frac{D_{T}T_{0}}{c^{2}}\delta\dot{\theta}=0\label{eq:4}\end{equation}
 Indeed, the two terms $-\frac{\kappa T_{0}}{c^{4}}\delta\ddot{\theta}$
and $-\frac{D_{T}T_{0}}{c^{2}}\delta\dot{\theta}$ in Eqs. (\ref{eq:3})
and (\ref{eq:4}) come from Eckart's proposal for the heat flux constitutive
equation depending on the hydrodynamic acceleration through the term
$-\frac{T}{c^{2}}\dot{u}^{\nu}$ (see Eq. (\ref{eq:fr})). All quantities
appearing in these equations are the same ones that appear in Eqs.
(\ref{eq:m1}-\ref{eq:m3}).

Proceeding as in the previous case, we analyze the dispersion relation
which is now given by

\begin{equation}
\left|\begin{array}{ccc}
s & n_{0} & 0\\
-\frac{1}{n\kappa_{T}}q^{2} & -\frac{\kappa T_{0}}{c^{4}}s^{2}+\tilde{\rho}_{0}s+Aq^{2} & \frac{\kappa}{c^{2}}q^{2}s-\frac{\beta}{\kappa_{T}}q^{2}\\
0 & \frac{T_{0}\beta}{n_{0}c_{n}\kappa_{T}}-\frac{D_{T}T_{0}}{c^{2}}s & s+D_{T}q^{2}\end{array}\right|=0\label{eq:8}\end{equation}
 which yields a quartic polynomial, namely\begin{equation}
b_{1}s^{4}+s^{3}+b_{2}s^{2}q^{2}+s\left(b_{3}q^{4}+a_{4}q^{2}\right)+b_{5}q^{4}=0\label{eq:9}\end{equation}
 with the coefficients given by\[
b_{1}=-\frac{\kappa T_{0}}{c^{4}\tilde{\rho}_{0}}\]
 \[
b_{2}=\frac{A}{\tilde{\rho}_{0}}+D_{T}\left(1-\frac{2\beta T_{0}}{c^{2}\kappa_{T}\tilde{\rho}_{0}}\right)\]
 \[
b_{3}=\frac{AD_{T}}{\tilde{\rho}_{0}}\]
 \[
b_{4}=\frac{\gamma}{\kappa_{T}\tilde{\rho}_{0}}\]
 \[
b_{5}=\frac{D_{T}}{\kappa_{T}\tilde{\rho}_{0}}\]
 We now attempt a rather intuitive but accurate solution to Eq. (\ref{eq:9}).
Notice that the coefficient of $s^{4}$ is very small. Neglecting
it, we can readily identify three roots, namely,\[
s_{1}=-\frac{b_{5}}{b_{4}}q^{2}\]
 \[
s_{2,3}=\left(-\frac{b_{2}}{2}+\frac{b_{5}}{2b_{4}}\right)q^{2}\pm i\sqrt{b_{4}}q\]
 Now, we assume that these three roots are still approximate solutions
to the quartic and find the fourth root by using the property that,
since the coefficient of $s^{3}$ is equal to one, $\sum_{i=1}^{4}s_{i}=-\frac{1}{b_{1}}$,
and thus\[
s_{4}\simeq\frac{c^{4}\tilde{\rho}_{0}}{\kappa T_{0}}+\left[\frac{A}{\tilde{\rho}_{0}}+D_{T}\left(1-\frac{2\beta T_{0}}{c^{2}\tilde{\rho}_{0}\kappa_{T}}\right)\right]q^{2}\]
 Since $\beta/\kappa_{T}<0$, the fourth root $s_{4}$ is always positive.
A real positive root in the dispersion relation yields an exponential
growth in the structure factor instead of a finite spectrum. This
behavior is unphysical and simply implies that the RB spectrum does
not exist in Eckart's formalism, even in the non-relativistic limit.

\section{modified eckart's theory}

The system of equations we consider in this section is obtained, as
in the previous one, from a momentum-energy tensor which includes
relativistic heat flux terms. The key difference here is that the
constitutive equation introduced for the heat flux is obtained through
the following argument. According to Eckart, such constitutive equation
is given by\begin{equation}
J_{[Q]}^{\nu}=-\kappa h_{\mu}^{\nu}\left(T^{,\mu}+\frac{T}{c^{2}}\dot{u}^{\mu}\right)\label{eq:fr}\end{equation}
 where the second term, as argued before \citep{Nos1} violates the
tenets of classical irreversible thermodynamics since it is neither
a thermodynamic force nor a flux. Further, it has another serious
drawback namely, it raises $\dot{u}^{\nu}$ to the category of a state
variable, a set already chosen to be given by $n$, $u^{\nu}$ and
$T$. Thus, the resulting set of hydrodynamic equations would be overdetermined.
Therefore, to keep Eq. (\ref{eq:fr}) to first order in the gradients,
we eliminate $\dot{u}^{\nu}$ using Euler's equation\begin{equation}
\tilde{\rho}_{0}\dot{u}^{\nu}=-p_{,\mu}h^{\mu\nu}\label{eq:fr1}\end{equation}
 Now, according to the local equilibrium assumption,\begin{equation}
p_{,\mu}=\frac{\beta}{\kappa_{T}}T_{,\mu}+\frac{1}{n_{0}\kappa_{T}}n_{,\mu}\label{eq:fr2}\end{equation}
 and thus, substitution of Eq. (\ref{eq:fr2}) in Eq. (\ref{eq:fr})
yields\begin{equation}
J_{[Q]}^{\ell}=-\kappa\left(1+\frac{T_{0}}{c^{2}\tilde{\rho}_{0}}\frac{\beta}{\kappa_{T}}\right)T^{,\ell}-\frac{\kappa T_{0}}{n_{0}\kappa_{T}c^{2}\tilde{\rho}_{0}}n^{,\ell}\label{eq:fr3}\end{equation}
 or\begin{equation}
J_{[Q]}^{\ell}=-L_{TT}T^{,\ell}-L_{nT}n^{,\ell}\label{eq:fr4}\end{equation}
 where $L_{TT}$ is an {}``effective thermal conductivity'' given
by\begin{equation}
L_{TT}=\kappa\left(1+\frac{\beta T_{0}}{c^{2}\tilde{\rho}_{0}\kappa_{T}}\right)\label{eq:hh}\end{equation}
 and $L_{nT}$ a new transport coefficient given by $\frac{\kappa T_{0}}{n_{0}\kappa_{T}c^{2}\tilde{\rho}_{0}}$
which has no classical counterpart. We would like to remark that an
equation similar in structure to Eq. (\ref{eq:fr4}) was already derived
by Landau and Lifshitz \citep{L&L}.

Clearly these transport coefficients can also be calculated from a
kinetic model but we shall discuss such calculation in a separate
paper. The equations in this formalism need not be justified in detail
since they follow simply from the method used to obtain Eckart's equations
incorporating the terms arising from Eq. (\ref{eq:fr4}) for the heat
flux. Thus we obtain that\begin{equation}
\delta\dot{n}+n_{0}\delta\theta=0\label{eq:e2}\end{equation}
 \begin{eqnarray}
\tilde{\rho}_{0}\delta\dot{\theta}+\frac{1}{n\kappa_{T}}\nabla^{2}\delta n+\frac{\beta}{\kappa_{T}}\nabla^{2}\delta T\nonumber \\
-A\nabla^{2}\delta\theta-\frac{L_{TT}}{c^{2}}\nabla^{2}\delta\dot{T}-\frac{L_{nT}}{c^{2}}\nabla^{2}\delta\dot{n} & =0\label{eq:e3}\end{eqnarray}
 \begin{equation}
\delta\dot{T}+\frac{T_{0}\beta}{n_{0}c_{n}\kappa_{T}}\delta\theta-\frac{L_{TT}}{n_{0}c_{n}}\nabla^{2}\delta T-\frac{L_{nT}}{n_{0}c_{n}}\nabla^{2}\delta n=0\label{eq:e4}\end{equation}
 and give rise to the dispersion relation,

\begin{equation}
\left|\begin{array}{ccc}
s & n_{0} & 0\\
-\frac{1}{n_{0}\kappa_{T}}q^{2}+\frac{L_{nT}}{c^{2}}sq^{2} & \tilde{\rho}_{0}s+Aq^{2} & \frac{L_{TT}}{c^{2}}q^{2}s-\frac{\beta}{\kappa_{T}}q^{2}\\
\frac{L_{nT}}{n_{0}c_{n}}q^{2} & \frac{T_{0}\beta}{n_{0}c_{n}\kappa_{T}} & s+\frac{L_{TT}}{n_{0}c_{n}}q^{2}\end{array}\right|=0\label{eq:8}\end{equation}
 which can be written as\begin{equation}
s^{3}+d_{2}s^{2}q^{2}+s\left(d_{3}q^{4}+d_{4}q^{2}\right)+d_{5}q^{4}\label{eq:fr5}\end{equation}
 where\begin{equation}
d_{2}=\frac{A}{\tilde{\rho}_{0}}+\frac{L_{TT}}{n_{0}c_{n}}\left(1-\frac{\beta T_{0}}{c^{2}\kappa_{T}\tilde{\rho}_{0}}\right)-\frac{n_{0}L_{nT}}{c^{2}\tilde{\rho}_{0}}\label{eq:fr6}\end{equation}
 \begin{equation}
d_{3}=\frac{AL_{TT}}{n_{0}c_{n}\tilde{\rho}_{0}}\label{eq:fr7}\end{equation}
 \begin{equation}
d_{4}=\frac{\gamma}{\kappa_{T}\tilde{\rho}_{0}}\label{eq:fr8}\end{equation}
 \begin{equation}
d_{5}=\frac{1}{\kappa_{T}\tilde{\rho}_{0}n_{0}c_{n}}\left(L_{TT}-\beta n_{0}L_{nT}\right)\label{eq:fr9}\end{equation}
 Once more, following the steps outlined for the two previous cases,
one can identify the modifications to the spectrum, as before, by
analyzing the roots. For this case the width of the Rayleigh peak
is given by\begin{equation}
\Delta_{RS}=\frac{d_{5}}{d_{4}}q^{2}=\frac{q^{2}}{n_{0}c_{n}\gamma}\left(L_{TT}-\beta n_{0}L_{nT}\right)\label{eq:fr10}\end{equation}
 The shift in the Brillouin doublet is the same as in the Meixner
case. In the particular case of an ideal gas, the properties of $\tilde{\rho}_{0}$
\citep{cer} guarantee that the position of the peaks will never exceed
$\pm cq$. On the other hand, the width is significantly modified.
We now obtain\begin{equation}
\Delta_{BS}=\frac{q^{2}}{2}\left\{ \frac{A}{\tilde{\rho}_{0}}+\frac{L_{TT}}{n_{0}c_{n}}\left(1-\frac{1}{\gamma}-\frac{\beta T_{0}}{c^{2}\kappa_{T}\tilde{\rho}_{0}}\right)+L_{nT}\left(\frac{\beta}{c_{n}\gamma}-\frac{n_{0}}{c^{2}\tilde{\rho}_{0}}\right)\right\} \label{eq:fr11}\end{equation}
 Equations (\ref{eq:fr10}) and (\ref{eq:fr11}) deserve further attention.
The former predicts a modification in Rayleigh's peak which changes
its width due to the effective thermal conductivity given by Eq. (\ref{eq:hh})
and the presence of $L_{nT}$; both, as stressed above, are strictly
relativistic effects. This is quite different from Meixner's like
theory where the correction arises only from the relativistic value
of $\kappa/c_{n}$. Moreover, the shape of Brillouin's peaks is further
altered due to several relativistic terms as can be seen from Eq.
(\ref{eq:fr11}).

\section{Summary and final remarks}

In the previous section, the modifications to the RB spectrum according
to the three versions of relativistic irreversible thermodynamics
have been explored. The difference between the modified Eckart's theory
result analyzed in Sect. IV and Meixner's case analyzed in Sect. II,
should be emphasized. The latter one does not have a density gradient
in {}``Fourier's equation'' which, as shown in Eq. (\ref{eq:fr3}),
is strictly a relativistic factor. This poses an intriguing question
namely, in both cases which are alternative versions of a relativistic
non-equilibrium theory, we predict the existence of a RB spectrum.
These spectra are different in both cases, but contain modifications
in comparison to the classical spectrum and both reduce to it in non-relativistic
limit. Which theory is the correct one? If we believe in fundamentals
we would be inclined to choose the version that is consistent with
the results obtained from kinetic theory and thus, as we have shown
in previous work \citep{physica A}, the second theory prevails. However,
Eckart's modified theory still contains the heat flux as a component
of the momentum-energy tensor which still is debatable. On the other
hand if we wish a phenomenological theory that contains a density
gradient in the heat flux within the Meixner's like formalism we would
face the problem of how to introduce such term. The answers to these
puzzles are still open. We feel that Eckart's original approach may
be discarded but the final answer as to which is the appropriate version
of a relativistic non-equilibrium thermodynamics is still a challenge
both theoretically and experimentally.

To facilitate the various results available for the RB spectrum we
are summarizing them in the Appendix.

\appendix

\section*{Appendix}

The modifications to the RB spectrum in the three frameworks considered
are summarized in the following table

\noindent \begin{tabular}{|c|c|c|c|}
\hline 
 & {\footnotesize Rayleigh's peak width} & {\footnotesize Brillouin peaks shift} & {\footnotesize Brillouin peaks width}\tabularnewline
\hline
\hline 
{\footnotesize Non-Relativistic} & {\footnotesize $\frac{D_{T}}{\gamma}q^{2}$} & {\footnotesize $\pm\sqrt{\frac{\gamma}{\rho_{0}\kappa_{T}}}q$} & {\footnotesize $\frac{1}{2}\left(\frac{A}{\rho_{0}}+\frac{\gamma-1}{\gamma}D_{T}\right)q^{2}$}\tabularnewline
\hline 
{\footnotesize Meixner} & {\footnotesize $\frac{\mathcal{D}_{T}}{\gamma}q^{2}$} & {\footnotesize $\pm\sqrt{\frac{\gamma}{\tilde{\rho}_{0}\kappa_{T}}}q$} & {\footnotesize $\frac{1}{2}\left(\frac{\mathcal{A}}{\tilde{\rho}_{0}}+\frac{\gamma-1}{\gamma}\mathcal{D}_{T}\right)q^{2}$}\tabularnewline
\hline 
{\footnotesize Eckart} & {\footnotesize No spectrum} & {\footnotesize No spectrum} & {\footnotesize No spectrum}\tabularnewline
\hline 
{\footnotesize Modified Eckart} & $\frac{q^{2}}{n_{0}c_{n}\gamma}\left(L_{TT}-\beta n_{0}L_{nT}\right)$ & {\footnotesize $\pm\sqrt{\frac{\gamma}{\tilde{\rho}_{0}\kappa_{T}}}q$} & {\scriptsize $\frac{q^{2}}{2}\left\{ \frac{A}{\tilde{\rho}_{0}}+\frac{L_{TT}}{n_{0}c_{n}}\left(\frac{\gamma-1}{\gamma}-\frac{\beta T_{0}}{c^{2}\kappa_{T}\tilde{\rho}_{0}}\right)L_{nT}\left(\frac{\beta}{c_{n}\gamma}-\frac{n_{0}}{c^{2}\tilde{\rho}_{0}}\right)\right\} $}\tabularnewline
\hline
\end{tabular}

\noindent where calligraphic fonts are being used for relativistic
transport coefficients in order to distinguish them from the non-relativistic
ones.

\end{document}